\documentclass[fleqn,twoside]{article}
\usepackage{espcrc2,psfig}

\usepackage{graphicx}
\usepackage[figuresright]{rotating}

\newcommand{\AmS}{{\protect\the\textfont2
  A\kern-.1667em\lower.5ex\hbox{M}\kern-.125emS}}

\title{A study of large field configurations in MC simulations}
\author{Li Li\address[MCSD]{Department of Physics and Astronomy, \\
The University of Iowa, \\
Iowa City, Iowa 52242, USA}
       and
        Y. Meurice\addressmark[MCSD] \thanks{This 
research was supported in part by the Department of Energy
under Contract No. FG02-91ER40664.}}

\begin{document}

\begin{abstract}
We discuss a new approach of scalar field theory where the small field
contributions are treated perturbatively and the large field configurations
(which are responsible for the asymptotic behavior of the perturbative series)
are neglected. 
In two Borel summable 
$\lambda \phi ^4$ problems 
improved perturbative series can be obtained 
by this procedure.
The modified series converge towards values exponentially close to the 
exact ones.
For $\lambda$ larger than some critical value,
the method outperforms Pad\'e's approximants and Borel summations. 
The method can also be used for series which are not Borel summable 
such as the double-well potential series and provide a perturbative 
approach of the instanton contribution.
Semi-classical methods can be used to calculate the modified
Feynman rules, estimate the error and optimize the field cutoff.
We discuss Monte Carlo simulations in one and two dimensions which 
support the hypothesis of dilution of large field configurations
used in these semi-classical calculations. 
We show that Monte Carlo methods can be used to calculate the
modified perturbative series.
 
\vspace{1pc}
\end{abstract}

\maketitle

In a recent publication \cite{convpert}, we have shown with three 
nontrivial $\phi^4$ problems 
that cutting off the integration range of $\phi$ at each site modifies
the large order behavior of these series. The modified series {\it converge}
toward values which are exponentially close to the exact ones.
The examples were the 
anharmonic oscillator and the Landau-Ginzburg hierarchical model which 
are Borel summable, but also the double-well which is not Borel summable, 
and in this case, the method
incorporates instanton effects. 
More generally, we are interested in developing 
approximate treatments where the  
small field configurations are treated perturbatively 
and the large field configurations semi-classically.

All the examples discussed above can be solved accurately with 
numerical methods \cite{anh,gam}. In the following, we discuss the
possibility of using the Monte Carlo method to perform such calculations.
We consider scalar models with one components at each site. We define the 
norm of a configuration $C$ as $|C|=Max_x\lbrace{|\phi_x|}\rbrace$.
The norm distributions of 
MC configurations for $\lambda \phi^4$ models in 1 and 2 dimensions
are shown in Fig. 1.
One sees that the distributions have rapidly falling tails and 
that the number of 
large norm configurations is exponentially suppressed.
One would thus expect that the error 
on correlation functions caused  by cutting-off the configurations
with $|C|>\phi_{max}$ has the generic form 
\begin{equation}
{\rm Error} \sim \phi _{max}^{A_1}{\rm e}^
{- A_3\phi_{max}^{A_2}}\ .
\end{equation}
In addition, the sites with large fields are clustered, and the
configuration has a typical shape 
in these regions as shown in Fig. 2.
This suggests that the large field configurations contribute dilutely to
the functional integral. This is just what is needed for a semi-classical
calculation.

The similarity of the (Borel summable) results 
found in Ref. \cite{convpert} suggests that in general 
the corrections due to the field cutoffs can be 
expressed as simple one-dimensional integrals. 
The correction to the zeroth order of the 
ground state of the anharmonic oscillator 
(in other words, the correction to the ground state of the harmonic oscillator)
has the semi-empirical form \cite{convpert}.
\begin{equation}
\delta E_0^{(0)} \simeq
4 \pi^{-1/2}\phi_{max}^{2}\int_{\phi_{max}}^{+\infty}d\phi{\rm e}^{-\phi^2}\ .
\nonumber
\end{equation}
\begin{figure}[htb]
\centerline{\psfig{figure=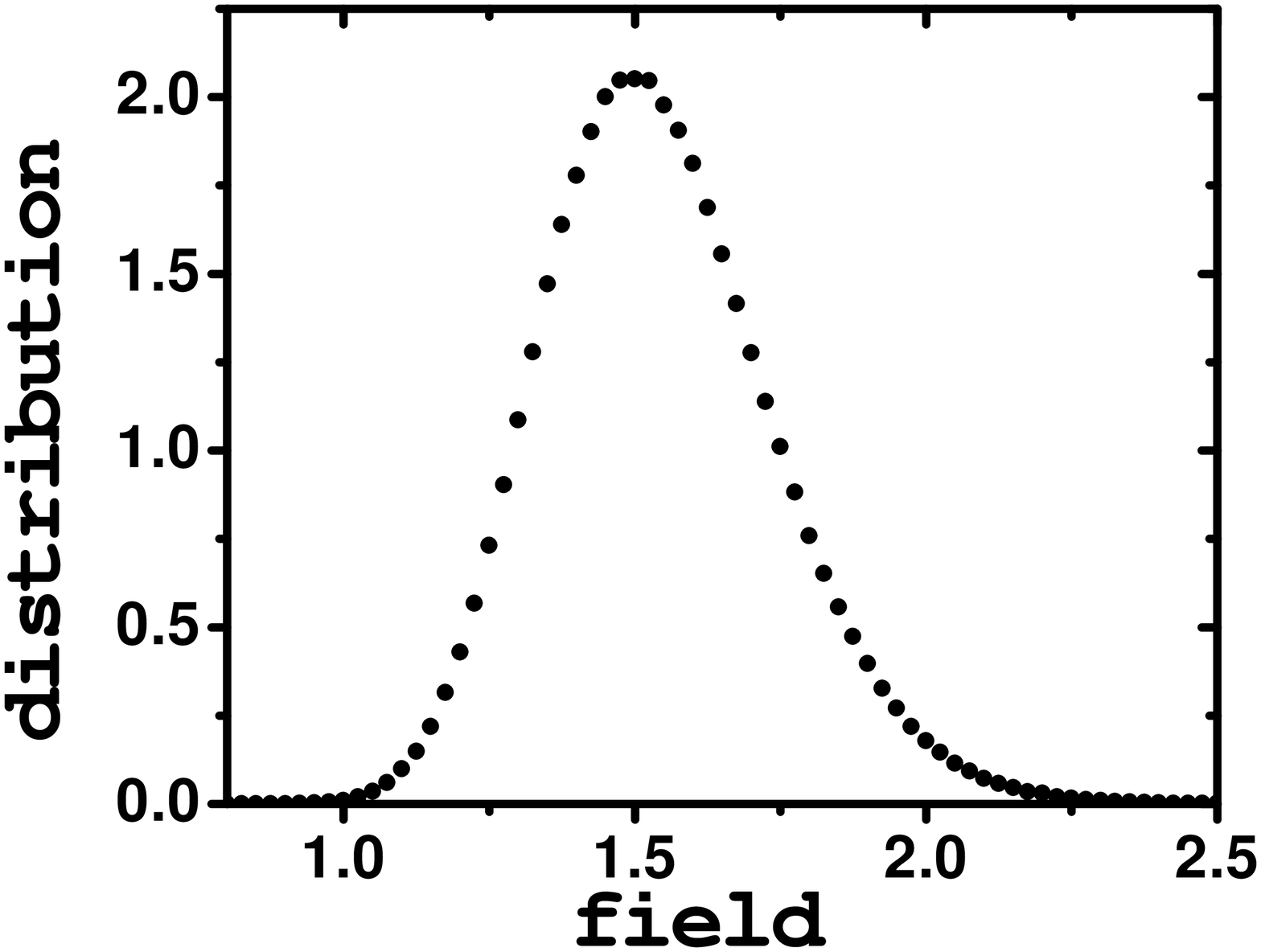,width=2.5in}}
\centerline{\psfig{figure=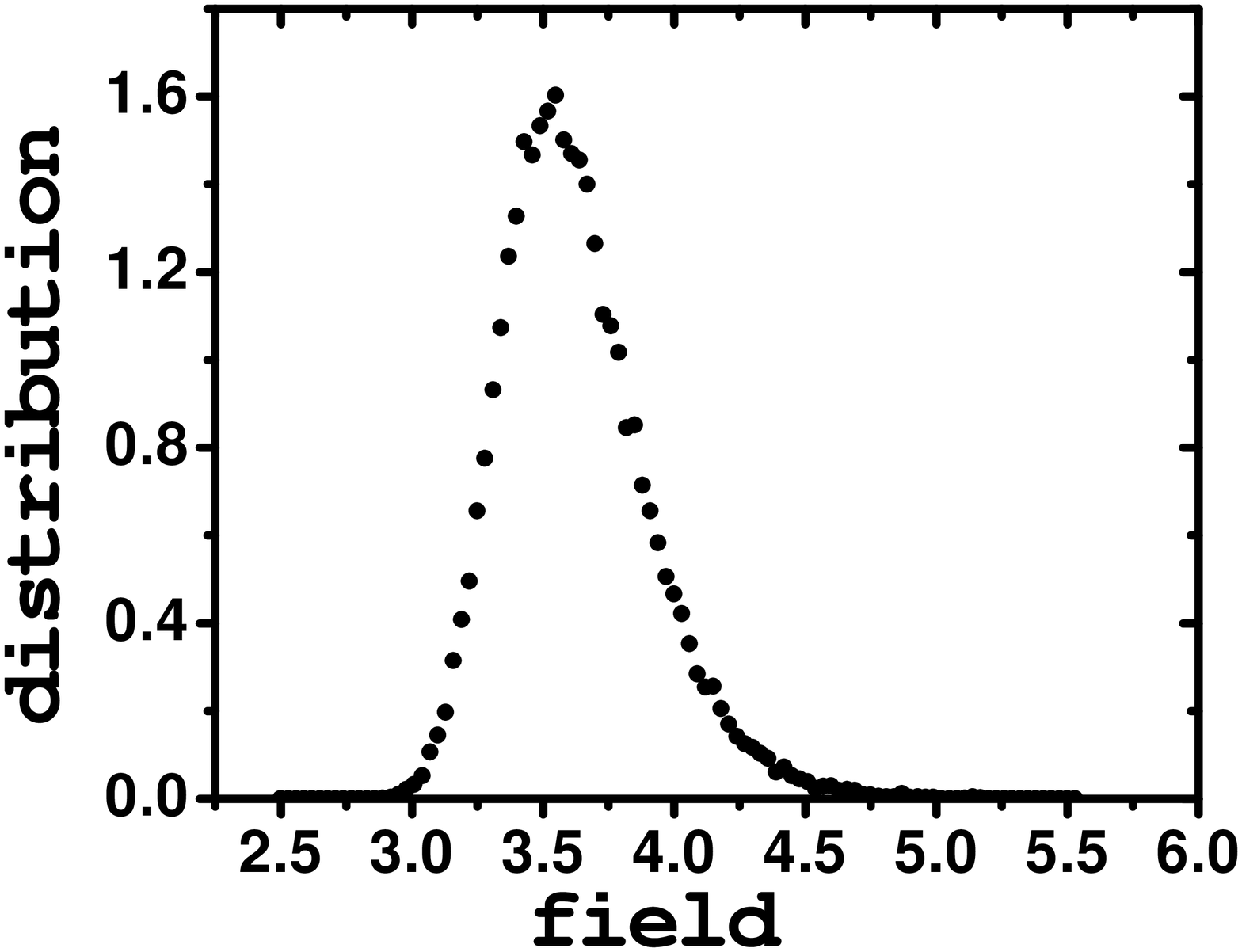,width=2.5in}}
\caption{Norm distribution for $\lambda \phi^4$ models in 1 and 2 dimensions}
\end{figure}
\begin{figure}[htb]
\vskip-3pt
\centerline{\psfig{figure=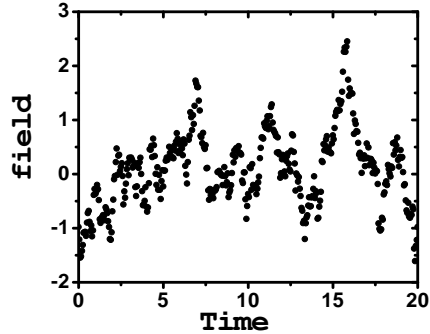,width=2.6in}}
\caption{A large norm configuration for the anharmonic oscillator}
\end{figure}
\begin{figure}[htb]
\centerline{\psfig{figure=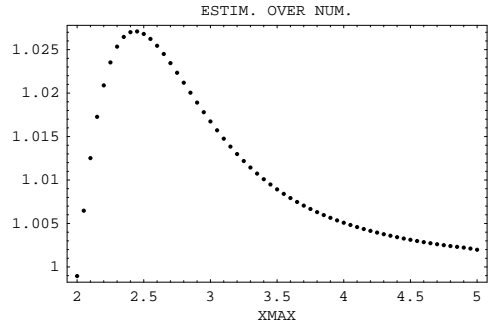,width=2.5in}}
\vskip-5pt
\caption{Ratio of the estimate and the numerical value for 
$\delta E_0^{(0)}$ vs $\phi_{max}$.}
\end{figure}
Fig. 3 shows that this is a good approximation which becomes better 
as $\phi_{max}$ increases.

When $\phi_{max}$ is large, the semi-empirical formula has the 
asymptotic form
\begin{equation}
\delta E_0^{(0)} \simeq
2 \pi^{-1/2}\phi_{max}{\rm e}^{-\phi_{max}^2}\ \nonumber
\end{equation}
\vskip10pt
This correction has the same functional dependence as the 
semi-classical splitting for the first two levels of the 
double-well \cite{coleman}:
\begin{equation}
\delta E_0^{(0)} \propto
\bigl({S_0\over {2\pi}}\bigr)^{1/2}{\rm e}^{-S_o}  \nonumber
\end{equation}
provided that we take $S_0=\phi_{max}^2$, the classical action 
from the classical configuration $\phi_{max} {\rm e}^{-|\tau-\tau_0|}$.
This confirms the validity of the dilute gas approximation.
Calculations of higher order perturbative coefficients and in 
higher dimensions are in progress.

It is possible to use the MC method to calculate the 
coefficients of the modified perturbative series. This will 
be particularly useful to check the validity of semi-classical 
methods in higher dimensions. In the case of the anharmonic oscillator 
$H=p^2/2+\phi^2/2+\lambda\phi^4$,
the first order coefficient is
\begin{equation}
<\sum_x \phi^4_x>_{\rm Gaussian}/\sum_x =3/4 \ . 
\end{equation}
We have calculated the 
modified averages corresponding to various field cuts.
The results are shown in Fig. 4 and are in good
agreement with the accurate numerical results. 
Note that these coefficients 
have a sizable lattice spacing dependence which can taken care 
with various methods which will be discussed elsewhere\cite{lili}.
\begin{figure}[htb]
\vskip-6pt
\centerline{\psfig{figure=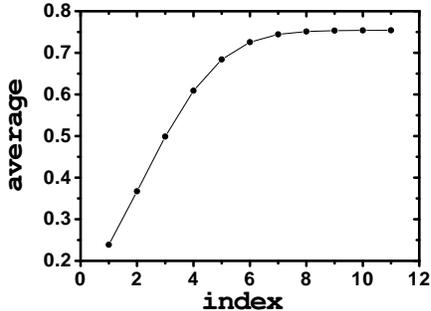,width=2.6in}}
\vskip-8pt
\caption{First perturbative coefficient for the ground 
state of the  anharmonic oscillator with
different field cuts : 1 is 1.5, 2 is 1.75, 3 is 2 etc. . 
The continuous line is
the numerical answer. }
\end{figure}

At a fixed value of $\phi_{max}$, the error due to the field cut
decreases when the coupling increases. At a fixed order in the 
modified perturbative series corresponding to the 
same fixed $\phi_{max}$, the  error due to the omission of the 
higher orders decreases when the coupling decreases. Consequently, 
at fixed $\phi_{max}$, there is a value of the coupling where the
the accuracy is optimal. This value is approximately where the 
two error curves meet. Conversely, at 
fixed coupling, one can find an optimal value of $\phi_{max}$.
This value can be related to 
the optimal values of variational parameters used in 
Refs. \cite{bdj,jk}. In these calculations, an intermediate 
mass term is introduced. The effect of this term is at the same 
time to reduce the contributions of the large fields and to reduce
the argument of the exponential which is expanded. 
Simple power law relations appear \cite{prep} 
in the strong and weak coupling 
limits for the problem of the one dimensional integral \cite{bdj}.

The coupling dependence of the optimal $\phi_{max}$ is 
a question which needs to be understood if we want to apply similar 
methods to gauge theories. In Wilson's formulation, 
at fixed lattice spacing, a field 
cut inversely proportional to the gauge coupling is 
imposed when one uses the compact group integration.
Consequently, the possibility for scalar models 
of having an optimal $\phi_{max}$ 
varying like the inverse square root of the quartic coupling
in some regime would be quite interesting.
We have started investigating the possibility of introducing 
field cuts in gauge theories by using the public OSU quenched 
configurations \cite{osu}. These configurations are in the Landau 
gauge, and the distributions of values of a fixed matrix element
have similarities with the scalar case.

Finally, we would like to mention that, in the large-$N$ limit, 
the critical potentials
of the $O(N)$ models in 3-dimensions have a finite radius 
of convergence when expanded in the $O(N)$ invariant 
quantity $\phi^2$ \cite{largen}. 
This is due to two conjugated branch cuts in 
the complex $\phi^2$ plane.
It is tempting to consider the model with a field cut 
which corresponds to this radius of convergence. However, this 
procedure generates sizable errors compared to the 
reliable procedure which consists in defining the critical potential 
using Pad\' e 
approximants.

\end{document}